\documentclass{iaus}
\usepackage{graphicx}

\title[Open cluster system] %% give here short title %%
{Observational properties of the open cluster system of the Milky Way and what they tell us about our Galaxy}

\author[A. Moitinho]   %% give here short author list %%
{Andr\'e Moitinho}
\affiliation{SIM/FCUL, Faculdade de Ci\^encias da Universidade de
              Lisboa, Ed. C8, Campo Grande, 1749-016 Lisbon, Portugal\\ email: {\tt andre@sim.ul.pt}}

\pubyear{2010}
\volume{266}  %% insert here IAU Symposium No.
\jname{Star clusters: basic galactic building blocks}
\editors{R. de Grijs \& J. R. D. L\'epine, eds.}
\begin{document}

\maketitle

\begin{abstract}
Almost 80 years have passed since Trumpler's analysis of the Galactic
open cluster system laid one of the main foundations for understanding
the nature and structure of the Milky Way. Since then, the open
cluster system has been recognised as a key source of information for
addressing a wide range of questions about the structure and evolution
of our Galaxy. Over the last decade, surveys and individual
observations from the ground and space have led to an explosion of
astrometric, kinematic and multiwavelength photometric and
spectroscopic open cluster data. In addition, a growing fraction of
these data is often time-resolved. Together with increasing computing
power and developments in classification techniques, the open cluster
system reveals an increasingly clearer and more complete picture of
our Galaxy. In this contribution, I review the observational
properties of the Milky Way's open cluster system. I discuss what they
can and cannot teach us now and in the near future about several
topics such as the Galaxy's spiral structure and dynamics, chemical
evolution, large-scale star formation, stellar populations and more.
\end{abstract}

\firstsection % if your document starts with a section,
\section{Preamble}

In 1930, Robert Trumpler \nocite{Trumpler1930} published his seminal
paper in which he proves the existence of the interstellar medium
(ISM).  In that paper, Trumpler showed how distances to open clusters
(OCs) derived from the apparent magnitudes of cluster stars of known
spectral types are systematically greater than those derived from
their apparent sizes, and how the effect increases with distance.  He
correctly attributed this phenomenon to the presence of an intervening
medium. This was the first paper using OCs as tracers of Galactic
structure.

\section{Overview}

Since the work of Trumpler, the open cluster system has been used to
characterise, study and analyse a variety of aspects related to the
structure, composition, dynamics, formation and evolution of the Milky
Way. Not being exhaustive, these aspects include:
\begin{itemize}
\item Spatial scales: scale height of disk(s), height of the Sun above
the Galactic plane, distance to the Galactic Centre.
\item Spiral structure: tracing spiral arms.
\item Kinematics and dynamics: rotation of spiral pattern, Galactic
rotation curve (velocity of the `local standard of rest', Oort
constants), orbits of OCs and their relation to cluster
survival/disruption.
\item Galactic formation and evolution: ages of the oldest clusters,
abundance pattern, inflow rate.
\item Distribution (absorption) and kind (extinction law) of interstellar matter.
\end{itemize}

The potential of using OCs for studying the Milky Way is well known
and sentences like ``open clusters are valuable objects for Galactic
studies'' appear in the introduction of almost every OC paper.
Leaving aside the astrophysical reasons, the main advantage of using
OCs is the precision with which one can derive their reddening values,
distances, ages, velocities and metallicities, compared to individual
stars. This arises not only from improved statistics, as OCs are
ensembles of stars, but also from the distribution of cluster stars in
photometric diagrams. In fact, the colour--magnitude and two-colour
diagrams (CMDs and TCDs, respectively) of star clusters can exhibit a
rich structure that provides bench marks to constrain those parameters
beyond a simple statistical average. An example is the `elbow'
displayed in optical CMDs for early $A$-type stars, which is very
useful for constraining distances.

There are also some disadvantages, or at least difficulties and
systematics in studying the Galaxy on the basis of OCs.  Confusion
against crowded backgrounds in the Galactic plane and dissolution of
OCs likely produce selection effects that conspire against the
identification of faint and/or poorly populated clusters. These
effects will produce a selection against old clusters, which are
fainter and potentially poorer (due to dynamical evolution) than their
younger counterparts, and against the low-mass end of the OC
population.  The extent of such a selection is far from well known,
but it must be kept in mind when deriving conclusions from the
observed global properties of the OC system.

Using OCs for studying the Galaxy relies on a huge amount of
painstaking data gathering.  Each cluster is just a point in parameter
space, but requires measuring many stars. Traditionally, stars were
measured one by one using photoelectric photometers and
spectrographs. Modern CCDs, infrared arrays and multi-object
spectrographs (MOSs) have the ability of observing hundreds or
thousands of stars in one shot, but reducing the data has proved to be
very time consuming. In the last few years, increased computing power
and development of automated or semi-automated reduction procedures
have alleviated the reduction burden.  But it took several years for
the community to master software like {\sc daophot} (Stetson et
al. 1990).  I guess that many of us have known what it is to reduce
images one by one, selecting by hand the adequate PSF stars, testing
the PSF model and redoing it if necessary. In the mid 1990s, the
workstation I had at work would take several hours just to run {\sc
allstar} on a $2048\times 2048$-pixel frame. Now, this can be done in
a few minutes, or less, and most of us have used experience to devise
automatic ways to select the proper PSF stars, so that a whole
observing run can be reduced fairly quickly.

The actual job of studying the Galaxy requires a catalogue of OC
parameters (reddening values, distances, ages, velocities,
metallicities, diameters). Thousands of studies of individual OCs
exist in the literature, often with nonconcordant results which must
be critically inspected and collated. Despite the care put into the
process, the variety of techniques used by different authors to
observe, reduce and derive parameters, lead to inhomogeneous
compilations that can limit their usefulness.  This problem is well
known and has motivated several attempts of deriving homogeneous
parameter sets, usually by adopting some well-defined analysis
procedure, reference curves (e.g., the zero-age main sequence, ZAMS,
and isochrones) and measurements. Still, the problem persists for most
objects.

It is therefore clear that assembling the observational basis for
studying the Galaxy is an incremental process. It is built up slowly
by adding the results of individual cluster studies, each one
contributing one probing point. It is therefore natural that
qualitative leaps in our understanding of the Galaxy happen in
intervals of many years. Roughly, such jumps take about 20 years.

Very schematically, in the 1980s there was the Lund catalogue
(Lyng{\aa} 1987), which contained around 1200 clusters of which a few
hundred had some of their parameters determined. The classic review on
OCs and Galactic structure is that of Janes \& Adler (1982) and its
update (Janes et al. 1988). The 1990s brought us the on-line WEBDA OC
database (formerly BDA; Mermilliod 1992) which not only included a
list of clusters and their parameters (essentially the Lund
catalogue), but also the actual measurements of their stars
(photometry, astrometry, spectral types, etc.). From the observational
point of view, CCDs and infrared arrays became of common use, {\sl
Hipparcos} provided a wealth of astrometric and kinematic data and
settled the distance scale through accurate parallaxes of the Hyades
and Pleiades. Although the new data did not bring in new significant
results regarding the spatial structure of the Milky Way, the number
of abundance measurements and discovery of more old OCs shifted the
focus on the origin and chemical evolution of the Galaxy. The
canonical review on the subject is that of Friel (1995) and its
updates.  The 2000s brought the Dias et al. (2002) catalogue of OCs,
which updates the Lund catalogue. The latest version includes 1787
clusters and has eliminated a number of entries shown to be
nonexistent. The {\sc 2mass} survey (Cutrie et al. 2003) has triggered
searches of new clusters (optically detectable and embedded, e.g.,
Bica et al. 2003b; Dutra et al. 2003; Froebrich et al. 2007) but very
few of them have actually been studied. So far, this decade has not
yet provided a fresh review on our subject, although some refined
recalculations taking advantage of currently available data have been
published (Bonatto et al. 2006; Piskunov et al. 2006).  However, the
rapidly growing number of spectroscopic observations allowed by MOSs,
availability of reduction pipelines and ongoing large-scale
photometric surveys are expected to soon take the understanding of our
Galaxy to the next step.

\section{The sample}

As discussed above, knowledge of OCs is compiled in catalogues. The
most complete catalogue of optically revealed OCs is that of Dias et
al. (2002)\footnote{http://www.astro.iag.usp.br/{\textasciitilde}wilton/}. Its
latest version (2.10) lists 1784 objects of which 969 (54\%) have
simultaneous reddening, distance and age determinations. The number of
clusters with distance, ages, radial velocities and proper motions is
much smaller (430, or 24\%) and only 178 (10\%) have their
metallicities determined. Lists of infrared (IR) OCs and candidates
have been compiled by Bica et al. (2003a,b), Dutra et al. (2003) and
Froebrich et al. (2007), based on searches using {\sc 2mass} data
products. These objects, detected in the IR, add about 700 clusters to
the optically identified family. However, very few have been studied
so far and many might not be real OCs. The main references for the IR
cluster lists are Bica et al. (2003a,b) and Dutra et al. (2003). The
systematic IR search of Froebrich et al. (2007) has produced a list of
around 1000 candidates, which they estimate suffers from 50\%
false-detection contamination. Once considering the number of already
identified IR clusters, the Froebrich et al. (2007) list still
contributes with the discovery of about 150 OCs (Glushkova et
al. 2009).

\begin{figure}[h]
\begin{center}
\includegraphics[width=12cm]{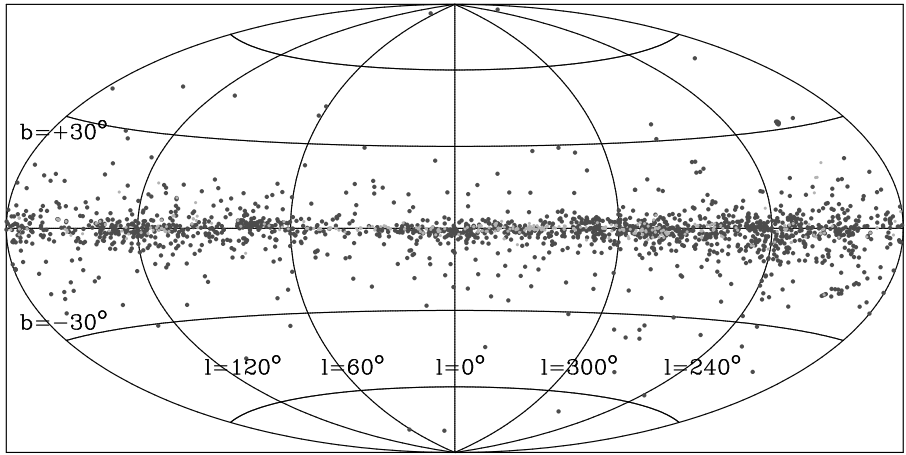} 
\caption{Aitoff distribution of known open clusters and candidates on
the Galactic plane. Black: optically visible clusters. Grey: infrared
clusters.}
   \label{fig:aitof}
\end{center}
\end{figure}

The distribution of optical and IR clusters is shown in Figure
\ref{fig:aitof}. It is noticeable how the IR clusters are mostly
located close to the $b=0^{\circ}$ plane, while the optical clusters
define a thicker disk. This is expected, since most obscuring material
is defining what is known as the extreme thin disk. The number
distribution of clusters is not uniform along the Galactic longitude
range. Moreover, clusters in different age groups exhibit different
distributions (e.g., figure~3 of Bonatto et al. 2006).

\begin{figure}[h]
% \vspace*{-2.0 cm}
\begin{center}
\includegraphics[width=9cm]{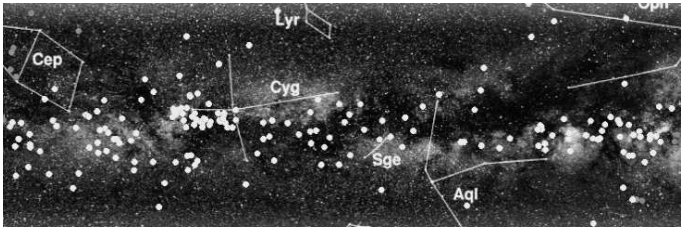} 
% \vspace*{-1.0 cm}
 \caption{Optically visible OCs (circles) plotted over an optical
 image of the first quadrant of the Milky Way (credit A. Mellinger).}
   \label{fig:extinzoom}
\end{center}
\end{figure}

At least two obvious biases modulate the distribution of
\emph{identified clusters}. The first is introduced by the extinction
produced by dense interstellar material. Figure \ref{fig:extinzoom}
clearly shows how the OCs delineate the wavy dust disposition. That
extinction affects the detectability of sources is well known, but the
kind of patterns it introduces in the distribution of OCs is often
overlooked in studies of the Galaxy using OCs (and other objects).
The other bias is produced by crowded backgrounds, against which an OC
might contribute little to the density of field stars and can thus
remain unidentified. This effect also includes nearby sparse OCs that
occupy large areas on the sky.

But there are also biases in the sample of \emph{studied clusters}. On
the one hand, interest has focussed on studying specific regions of
the Galaxy (e.g., Cygnus, Carina and Puppis--Vela), thus artificially
increasing the number of studied clusters in those regions with
respect to others. On the other hand, many clusters have been observed
for reasons other than studying the Galaxy. This was especially true
in the past, when focus was mostly on studying stellar evolution.  The
tendency has been to study bright and/or well-populated clusters,
which would provide much clearer Hertzsprung--Russell diagrams. Given
the aforementioned difficulties in data reduction, there was little
motivation to study poorer (potentially older) and fainter
(potentially older/more distant/more reddened).

Although these biases exist in principle, all-sky proper-motion and
photometric surveys can help to overcome them. The {\sl
Hipparcos}/Tycho-2 (ESA 1997; H{\o}g et al. 2000) and {\sc ucac2}
(Zacharias et al. 2004) astrometric (and photometric) catalogues, the
{\sc 2mass} $JHK$ photometry and images, as well as other available
data have enabled identification of OCs beyond simple visual
recognition. Searches for concentrations of stars at a common distance
or with common proper motion, cluster sequences in CMDs or simply
spatial concentrations of stars obeying some photometric criteria have
allowed discovery of more than a hundred optically visible OCs (e.g.,
Platais et al. 1998; Alessi et al. 2003; Kharchenko et al. 2005) and a
few hundred IR clusters, as mentioned above.

\begin{figure}[h]
\begin{center}
 \includegraphics[height=8cm]{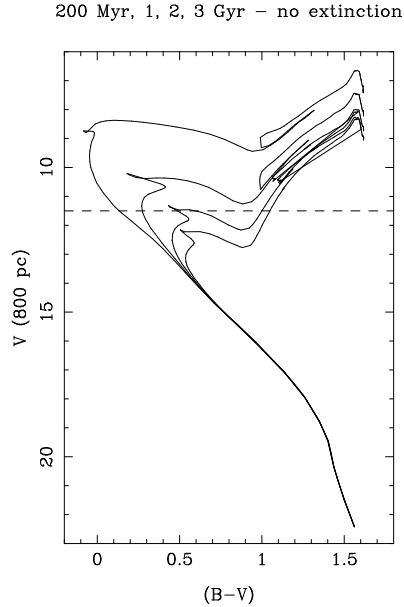} 
\caption{0.2, 1, 2 and 3 Gyr isochrones (Girardi et al. 2000) at a
distance of 800 pc from the Sun. No extinction is applied. The dashed
line indicates the 11.5 mag completeness limit of the stellar data
catalogue used in Piskunov et al. (2006).}
   \label{fig:complete}
\end{center}
\end{figure}

It has been argued that the sample of known OCs is complete within 850
pc (Piskunov et al. 2006; 1 kpc is also quoted). The argument is based
on the fact that the number of clusters kpc$^{-1}$ starts to decrease
beyond 850 pc.  It seems, however, that even within 850 pc the sample
of OCs might be missing a significant number of poorly populated
intermediate-age and old objects. Figure \ref{fig:complete} shows a
series of isochrones, from moderately young to moderately old, placed
at 800 pc from the Sun, with no extinction applied. It is readily seen
that at 1 Gyr and older, only the evolved portion of the cluster
sequence is above the dashed line, which marks the completeness limit
of the astrometric and photometric catalogue used in Piskunov et
al. (2006). Apart from the red clump, the other post-main-sequence
regions of the CMD will be barely populated. The red clump will only
be noticeable for fairly populated clusters. For the not-so-rich
clusters, the number of clump stars might not be enough to produce a
visible cluster signature either in proper-motion vector-point
diagrams or in CMDs. The ability to identify a cluster based on
kinematics will also be hampered along directions dominated by the
solar motion. Furthermore, in reality, clusters will be affected by
extinction, which has the effect of making the sequences fainter, thus
pushing them even further below the completeness line. For the 200 Myr
example, the red clump is very sparsely populated and the presence of
1 mag of absorption will leave out the main sequence, thus making
it very hard to detect (unless it is visually obvious on a sky chart).

As for the total number of OCs in the Milky Way, estimates depend on
many hypotheses, such as how well the number of clusters per volume in
the solar neighbourhood represents its position relative to the
Galactic Centre, the density-decay law and cluster completeness. In
any case, estimates indicate numbers on the order of $10^5$ OCs
(Bonatto et al. 2006; Piskunov et al. 2006).

\section{Spatial scales}

We now proceed to see what the statistical properties of the OC
distribution tell us about the spatial scales of the Galactic disk and
our position within it.  The increased samples of OCs analysed in
recent studies essentially recover the classic results from the 1980s,
although in a clearer way. In particular, it is found that the disk
scale height (SH) defined by the young clusters is smaller than that
derived from older clusters, with the young and intermediate-age OCs
($< 200$ Myr versus $< 1$ Gyr) having SHs of 48 pc and $\sim 150$ pc,
respectively. For older clusters, the vertical distribution is
essentially uniform, so that no SH can be derived. Also, the
distribution of OCs younger than $\sim 1$ Gyr clearly shows how the
disk outside the solar circle is $1.4-2.0 \times$ wider than inside
this radius (Bonatto et al. 2006).  Assuming that OCs are
symmetrically distributed above and below the plane, the zero point of
the height distribution gives a measure of the distance of the Sun
from the Galactic plane. The current values determined from OCs range
from about 15 to 22 pc (Bonatto et al. 2006; Piskunov et al. 2006;
Joshi 2007).

These seem to be well-established properties of the OC system that
persist as the sample grows.  Interpretation is, however, not so
straightforward and unambiguous. One of the first things that call our
attention is that the SH defined by thin-disk stars ($\sim 200$ pc) is
much larger that that defined by the young OCs, although it is more or
less compatible with that of intermediate-age OCs. This could be seen
as a disk evolutionary effect, by which the disk would thicken with
age due to some heating mechanism such as collisions with external
systems or by different conditions in the ancient disk (Janes \&
Phelps 1994). However, it could also be an effect of the selective
destruction of OCs at lower heights. Indeed, the range of heights of
the different OC age groups are very similar. What is different is the
proportion of clusters at small to large heights (see, e.g., figure 7
of Bonatto et al. 2006). This might as well be simply reflecting the
destruction of clusters that tend to be close to the plane, which
would trim the distribution peak, leading to a flatter distribution
and hence a larger SH. This scenario seems consistent with the
apparent flatness of the old OC distribution, which is not seen in the
smaller sample of Janes \& Phelps (1994), who find a SH of $\sim 375$
pc. Verification requires a kinematic analysis.

From the perspective of OCs as progenitors of the field population,
the link between the OC and field-star SHs remains to be firmly
established. Kroupa (2002) has shown how (kinematically hot) stars
escaping star clusters, mostly while emerging from their parental
molecular clouds, can thicken Galactic disks and even produce thick
disks. This mechanism could provide the link. Another interesting
consequence is that if this mechanism were the sole avenue for
producing a thick disk, then we would not expect to find thick-disk
OCs. Finding these clusters would mean that other processes (such as
capture from an accreted galaxy) would (also) operate.

As for the thicker SH outside the solar circle, the result is
derived from a global comparison of the SH of outer versus inner
clusters. Its precise meaning is not straightforward and is most
likely a combination of effects such as the flaring of the disk and
the amplitude of the warp. Isolating both effects requires larger
samples of OCs to analyse the SH at smaller Galactic longitude
intervals.

OCs have also been used for estimating the distance to the Galactic
Centre. From a kinematic analysis of 301 OCs within 3 kpc from the
Sun, Shen \& Zhu (2007) derive $R_{\rm GC} \sim 7.95$ kpc, which is
consistent with the value the same authors derive from OB stars
(8.25 kpc). Although based on a simple Oort model, these
determinations agree with the well-known results of Genzel et
al. (2000; 7.8--8.2 kpc) from analysis of radial velocities of more
than 100 stars around the black hole Sgr A*, and of Eisenhauer et
al. (2003), who found $R_{\rm GC} = 7.94 \pm 0.42$ kpc based on
astrometric and radial-velocity observations of star S2 around Sgr A*.
These results support the long-known indications that the standard IAU
value, $R_{\rm GC} \equiv 8.5$ kpc, should be revised and lowered by
$\sim 500$ pc. A synthesis of the determinations of $R_{\rm GC}$ in
the past decades is shown in figure~2 of Shen \& Zhu (2007).

Another aspect concerning the spatial scales of the Galactic disk is
the extent of the warp.  By looking along low-extinction windows in
the third Galactic quadrant (3GQ), V\'azquez et al. (2008) have been
able use OCs to trace the disk out to Galactocentric distances greater
than 16 kpc. In their figure~6, one can see how the southern warp
reaches distances greater than 1 kpc below the formal Galactic plane.
Finally, Carraro et al. (2009) reported OCs at large Galactocentric
distances, with VdB--Hagen~4 reaching 20 kpc. This value is
significantly larger than the $\sim 14$ kpc truncation limit usually
adopted and sets a lower limit to the extent of the stellar disk as
probed by OCs.

\section{Spiral structure}

Young OCs have long been used for tracing spiral structure (e.g.,
Becker \& Fenkart 1970; Moffat et al. 1979). Since star formation in
the Milky Way occurs mainly in the spiral arms, it is expected that
the younger clusters, which have not yet departed significantly from
their birth sites, will trace the local spiral structure.  Compared to
gaseous spiral indicators such as H{\sc i}, H{\sc ii} and CO, which
are observed at radio wavelengths and for which we rely on rotation
curves to derive their distances, OC distances are much better
determined. Furthermore, it is well known that the velocity field in
spiral arms does not correspond exactly to that implied by the general
rotation curve. Indeed, spiral arms appear as glitches in empirical
rotation curves. Thus, one arrives at the situation of tracing spiral
structure using velocity--distance relations that do not apply well to
the regions where they are used. Despite this source of uncertainty,
radio observations roughly reveal the general spiral structure of the
Milky Way, with the advantage of being insensitive to obscuration and
thus being able to reach most of the Galaxy. OC distances are more
precise, but are mainly obtained from observations in the optical
domain and cannot probe beyond a few kpc due to interstellar
obscuration (except for a few absorption windows at certain Galactic
longitudes).

When mentioning young OCs as spiral tracers, the question arises as to
what is meant by young. Dias \& L\'epine (2005), in their figure~1,
show that OCs up to $\sim 12$ Myr remain in their parent arms. At
$\sim 20$ Myr, they are already drifting away from the arms, joining
the general disk population, although spiral traces are still
identifiable.  This figure is illustrative of spiral structure as
traced by OCs and also shown in many other studies. In particular, the
Carina--Sagittarius, local (or Orion) and Perseus arms are prominent
in the solar vicinity up to distances of $\sim 4-5$ kpc.

Expanding on this classical view, the recent papers of Carraro et
al. (2005), Moitinho et al. (2006) and V\'azquez et al. (2008) have
revealed the spiral structure in the 3GQ based on OCs, CO clouds and
early-type field stars.  These studies support a four-armed pattern
and have shown that the Outer (Norma--Cygnus) arm crosses the whole
3GQ, reaching distances beyond 10 kpc from the Sun and that the local
(Orion) arm is more than a simple spur, reaching the Outer arm. As for
Perseus, it is not traced by young OCs in this region, being
apparently disrupted by the crossing of the local arm, but is then
again traced by OCs in the 4GQ. An aspect of these analyses that is
worthwhile emphasising is the treatment of early-type field stars.  At
sufficiently large distances, early-type stars belonging to spiral
arms can be considered as spatially confined along their lines of
sight. These stars will thus produce cluster-like sequences in
photometric diagrams that allow determining average reddening values
and distances using the same techniques as used for analysing
clusters. Interestingly, these sequences had previously been
identified and baptised as `blue plumes' (because of their plume-like
appearance on the blue side of CMDs) and were interpreted as the
signature of a dwarf galaxy cannibalised by the Milky Way (the Canis
Major dwarf galaxy; Mart\'{\i}nez--Delgado et al. 2005).  However,
these blue plumes have been shown not only to trace the spiral
structure in the 3GQ, but also across the 2GQ (Carraro et al. 2008),
where due to superposition with evolved OCs they have often been
interpreted as blue stragglers.

\section{Kinematics and dynamics}

The 1990s and 2000s witnessed an explosion of astrometric data. The
main players being the {\sl Hipparcos} and Tycho-2 catalogues (ESA
1997; H{\o}g et al. 2000) and the {\sc ucac2} compilation (Zacharias
et al. 2004).  The new {\sc ucac3} catalogue is being released during
this IAU General Assembly.

These data have led to many studies of OCs. From the individual
cluster point of view, they have allowed to determine cluster
membership probabilities, cluster proper motions, internal dynamics
and to firmly set the Hyades and Pleiades luminosity scales, which are
crucial for calibrating the cosmological distance ladder and the ages
provided by stellar evolution models.  These astrometric data have
also been widely used for studying star-forming regions (which are out
of the scope of this review) containing OCs. But regarding large-scale
Galactic structure, they have not triggered great qualitative
progress. The main reason for this is the very limited distance range,
of a few hundred pc, that they can probe.

OCs have often been used for studying the local behaviour of the
Galactic rotation curve. However, a proper rotation curve covering a
sizable range of the Galaxy is still lacking. Work in progress in this
area was recently reported by Frinchaboy \& Majewski (2008).  The
outcome of most kinematic research has been to improve the
determination of parameters such as the local standard of rest
rotation speed, Oort's constants and the velocity of the Sun. An
example of such an analysis is summarised in table~2 of Piskunov et
al. (2006).  These improvements do not only come from the available
high-quality astrometric data, but also from the quantity of data,
which allows to resolve local structures such as star-formation
complexes and stellar streams and eliminate the systematics they
introduce (e.g., Piskunov et al. 2006).

In addition to these more classical calculations, OCs have also been
used recently to determine the rotation speed of the spiral
pattern. Dias \& L\'epine (2005) confirmed that spiral arms rotate as
rigid bodies, and determined the pattern speed by direct integration
of cluster orbits and by comparing the positions of clusters in
different age groups, leading to an adopted value of 24 km s$^{-1}$
kpc$^{-1}$. As a an additional result, they find that the co-rotation
radius is $1.06\pm 0.08 \times$ the solar Galactocentric distance. It
is also interesting to note that Bobylev et al. (2007) find a
different speed for the Orion arm compared to Carina--Sagittarius and
Perseus.  Another basic parameter is the epicyclic frequency (the
residual circular velocity after subtracting the Galactic rotation
curve) which L\'epine et al. (2008) find to be $42 \pm 4$ km s$^{-1}$
kpc$^{-1}$ at the solar radius.

\section{Formation and evolution}

The ages of the oldest OCs place a lower limit on the age of the
Galactic disk. Presently, the oldest-known clusters are (still)
NGC~6791 and Be~17 which have ages of around 10 Gyr (Carraro et
al. 1999; Salaris et al. 2004), which is younger than the halo,
although they almost fall in the range of the youngest globular
clusters.

Few old open clusters have been identified in the inner Galaxy. This
is usually attributed to the harsher conditions that would destroy OCs
at small Galactocentric radii. Curiously, the orbit of NGC~6791, which
is one of the oldest known OCs and is presently outside the solar
circle, can reach a distance of only 4.6 kpc from the Galactic Centre.

The most discussed aspects of Galactic evolution scenarios are those
related to chemical-abundance gradients.  Most observational studies
in the last two decades find a smoothly decreasing
metallicity--Galactic radius gradient. Recent studies determine a
gradient of $\sim -0.06$ dex kpc$^{-1}$ (Friel et al. 2002; Chen et
al. 2003). However, controversy has been generated by the work of
Twarog et al. (1997), who reexamined the metallicity scales of
published observations and found the radial metallicity distribution
to be better described by a step function with the transition
occurring at around 10 kpc from the Galactic Centre. The inner region
would be characterised by a shallow gradient (with an average [Fe/H] =
0.0 dex), followed by a $-0.3$ dex step and then an almost flat
metallicity distribution. On the theoretical side, models usually
yield simple gradients, although the recent paper of Magrini et
al. (2009) is able to predict a sudden change at around 7 kpc,
followed by an almost constant plateau starting at around 11 kpc.
From a dynamical point of view, the simulations of Corder \& Twarog
(2001) show how orbital diffusion can smooth the discontinuity
(assuming it is real) and how using younger clusters enhances the
detection probability of the step.  Another perspective is offered by
Mishurov et al. (2002) and L\'epine et al. (2003) who show how the
co-rotation resonance can break a smooth abundance gradient.

Regarding a possible age--metallicity relation, most studies fail to
find such a dependence over a cluster age range of almost 10 Gyr,
although there is marginal evidence from clusters at Galactocentric
radii larger that 10 kpc (see Friel 1995; Friel et al. 2002). The most
natural explanation remains that chemical enrichment must have
essentially taken place early in the Galaxy's evolution.

For a more in-depth view of OCs as probes of the origin and chemical
evolution of the Milky Way the reader is referred to the reviews of
Janes \& Phelps (1994), Friel (1995) and Friel et al. (2002).

\section{Interstellar extinction}

We now briefly return to the point from which this review started:
extinction.  Joshi (2005) studied the interstellar-absorption pattern
within $\sim 4-5$ kpc using catalogued reddening values towards 722
OCs. He found that absorption approximately follows a sinusoidal curve
with Galactic longitude, with the highest values occurring at $l \sim
30$--50$^\circ$ and the lowest values around $l \sim
220$--250$^\circ$.  Extinction is generally low in the 3GQ, increasing
only $\sim 0.07$ mag kpc$^{-1}$ between $160^\circ < l < 280^\circ$,
in agreement with the analysis by V\'azquez et al. (2008). Except for
the opaque barrier at $l \sim 265^\circ$ due to the Vela molecular
ridge, the low-absorption window continues in the 4GQ towards
Carina. In general, for the remaining longitude range, absorption is
higher and happens mostly between 1 and 2 kpc from the Sun.

Joshi (2005) also determined that the dust plane is tilted by
$0.6^\circ$ with respect to the $b=0^\circ$ plane. The maximum and
minimum height of the dust plane occur at $\sim 54$ and $234^\circ$,
respectively, coincident with the maximum and minimum of the Galactic
warp.

Although these and other studies reveal the distribution and amount of
interstellar matter, they do not give any idea of the nature of the
absorbing material. This is due to most OC reddening and absorption
values having been derived assuming a standard interstellar extinction
law. A study of the variations of the interstellar law would certainly
be interesting.

\section{Prospects}

As insistently advocated in this contribution, an important goal for
the future is to have an unbiased, or at least well-understood, sample
of OCs. This will require mitigating extinction artifacts, which
demands analysing the mostly unstudied infrared clusters. On the way,
we will learn much about the kind, quantity and distribution of
interstellar material. Success will, however, depend on the
development of IR photometric analysis techniques capable of yielding
precise cluster parameters.

Another well-known problem to be addressed is that of the homogeneity
of OC-parameter compilations. Homogeneity of the data itself is also
certainly an issue, but of second order when compared to the
analysis. However, a word of caution should be issued: homogeneous
methods do not necessarily produce homogeneous catalogues. As an
example, {\sl Hipparcos} parallaxes are great for determining the
distance to close clusters such as the Pleiades. For a cluster at 300
pc, parallaxes will still be available but will not do a very good
job. In this case, a good ZAMS fit would produce a distance on a scale
in better agreement with the Pleiades distance from parallaxes.

As for the future, one cannot emphasise enough how important {\sl
Gaia} will be to Galactic astronomy. {\sl Gaia} will deliver excellent
distance estimates and kinematics for single stars, but it will not
end the use of OCs for studying the Galaxy. Clusters are still unique
when we need ages, and {\sl Gaia} will help improving them. Also, {\sl
Gaia} is not an IR mission and will therefore not be able to penetrate
dense regions.

Many large-area surveys are either ongoing or planned for the near
future using ground-based facilities: {\sc ukidss, segue, iphas,
vphas+, vvv, vhs, lsst, Pan-STARRS} and others. These surveys will
produce a wealth of data from the optical to the near-IR, including in
selected bands such as H$\alpha$. In several cases, observations will
even be time-resolved.

The next few years promise to offer a profusion of intense emotions to
OC researchers.

\begin{acknowledgements}
Although I appear as the sole author of this contribution, it has also
been the product of years of stimulating discussions with several
colleagues. I am particularly grateful to Wilton Dias, Giovanni
Carraro, Rub\'en V\'azquez and Jacques L\'epine.  This work was
partially supported under a bilateral FCT/CAPES agreement.
\end{acknowledgements}

\end{document}